\documentclass[conference]{IEEEtran}
\usepackage{amsmath,amsfonts}
\usepackage{algorithmic}
\usepackage{algorithm}
\usepackage{array}
\usepackage[caption=false,font=normalsize,labelfont=sf,textfont=sf]{subfig}
\usepackage{textcomp}
\usepackage{stfloats}
\usepackage{url}
\usepackage{verbatim}
\usepackage{graphicx}
\usepackage{cite}
\usepackage{braket}
\usepackage{soul}
\usepackage{booktabs}
\usepackage[most]{tcolorbox}
\usepackage{xcolor}
\definecolor{myhighlightcolor}{RGB}{255,255,153} 
\sethlcolor{myhighlightcolor} 

\definecolor{rqbg}{RGB}{232,238,255} 

\newtcolorbox{rqsummarybox}{
  colback=rqbg,
  colframe=rqbg,   
  boxrule=0pt,
  arc=0pt,         
  boxsep=0pt,
  left=5pt,right=5pt,top=4pt,bottom=4pt,
  width=\linewidth,
  before skip=6pt,
  after skip=6pt,
}

\def\BibTeX{{\rm B\kern-.05em{\sc i\kern-.025em b}\kern-.08em
    T\kern-.1667em\lower.7ex\hbox{E}\kern-.125emX}}
\begin{document}


\title{A Graph-Based Forensic Framework for Inferring Hardware Noise of Cloud Quantum Backend}
\author{\IEEEauthorblockN{Subrata Das}
\IEEEauthorblockA{\textit{Department of Electrical Engineering} \\
\textit{Pennsylvania State University}, \\
State College, PA, USA \\
sjd6366@psu.edu}
\and

\IEEEauthorblockN{Archisman Ghosh}
\IEEEauthorblockA{\textit{Dept. of CSE} \\
\textit{Pennsylvania State University}, \\
State College, PA, USA \\
apg6127@psu.edu}

\and 
\IEEEauthorblockN{Swaroop Ghosh}
\IEEEauthorblockA{\textit{Dept. of EECS} \\
\textit{Pennsylvania State University}, \\
State College, PA, USA \\
szg212@psu.edu}}




\maketitle

\begin{abstract}

Cloud quantum platforms give users access to many backends with different qubit technologies, coupling layouts, and noise levels. The execution of a circuit, however, depends on internal allocation and routing policies that are not observable to the user. A provider may redirect jobs to more error-prone regions to conserve resources, balance load or for other opaque reasons, causing degradation in fidelity while still presenting stale or averaged calibration data. This lack of transparency creates a security gap: users cannot verify whether their circuits were executed on the hardware for which they were charged. Forensic methods that infer backend behavior from user-visible artifacts are therefore becoming essential. In this work, we introduce a Graph Neural Network (GNN)-based forensic framework that predicts per-qubit and per-qubit link error rates of an unseen backend using only topology information and aggregated features extracted from transpiled circuits. We construct a dataset from several IBM 27-qubit devices, merge static calibration features with dynamic transpilation features and train separate GNN regressors for one- and two-qubit errors. At inference time, the model operates without access to calibration data from the target backend and reconstructs a complete error map from the features available to the user. This enables independent auditing by users, allows hardware providers to verify error data reported by third-party calibration services, and supports competitive analysis of new or undisclosed quantum hardware by peer providers. Our results on the target backend show accurate recovery of backend error rate, with an average mismatch of approximately 22\% for single-qubit errors and 18\% for qubit-link errors. These results are very good, given the error rates are in the order of $10^{-4}$ (single qubit gate error) to $10^{-2}$ (two qubit gate error). The model also exhibits strong ranking agreement, with the ordering induced by predicted error values closely matching that of the actual calibration errors, as reflected by high Spearman correlation. The framework consistently identifies weak links and high-noise qubits and remains robust under realistic temporal noise drift. This demonstrates that transpiled circuit statistics carry enough signal to support backend auditing and strengthen trust in cloud quantum execution environments.

\end{abstract}


\section{Introduction}
\label{sec:intro}
Cloud quantum platforms have made programmable quantum hardware widely accessible in the Noisy Intermediate-Scale Quantum (NISQ) era \cite{ibm_quantum_platform, rigetti_qcs_docs, ionq_website, debnath2016demonstration}. Users can choose among devices with different qubit technologies, gate sets, noise levels, and coupling architectures offered by third-party cloud providers. These services reduce the cost of experimentation, enable remote execution, and allow rapid prototyping of quantum algorithms. However, the increased reliance on third-party cloud systems brings forward a growing security concern: users are forced to trust hardware and software stacks that they cannot inspect \cite{ghosh_RE}.

Most providers expose only high-level interfaces to submit circuits and retrieve results. The internal routing, scheduling, and hardware allocation steps remain hidden. A backend may have multiple isomorphic subgraphs or several configurations with similar connectivity but different noise characteristics \cite{zulehner2018efficient,murali2019noise}. The transpiler is free to choose any of these during mapping and optimization. Even when a user specifies a target backend, the actual set of physical qubits used to execute the circuit is not guaranteed to match the public calibration data. Prior studies have shown that routing heuristics, layout selection, and dynamic resource contention can redirect circuits to noisier regions of the device \cite{murali2020software, larose2019overview}.

\begin{table}[t]
\centering
\caption{Fleet-level error summaries for IBM Heron QPUs (Accessed on 2025-12-15).}
\label{tab:ibm_heron_error_summary}
\begin{tabular}{cccc}
\hline
QPU & Median 2Q & Layered 2Q & Median Readout \\
\hline
ibm\_boston     & $1.25\times10^{-3}$ & $2.39\times10^{-3}$ & $4.395\times10^{-3}$ \\
ibm\_kingston   & $1.93\times10^{-3}$ & $3.13\times10^{-3}$ & $9.338\times10^{-3}$ \\
ibm\_pittsburgh & $1.74\times10^{-3}$ & $3.53\times10^{-3}$ & $4.395\times10^{-3}$ \\
ibm\_fez        & $2.64\times10^{-3}$ & $4.97\times10^{-3}$ & $9.766\times10^{-3}$ \\
ibm\_marrakesh  & $2.57\times10^{-3}$ & $4.04\times10^{-3}$ & $1.16\times10^{-2}$ \\
\hline
\end{tabular}
\end{table}

The above lack of transparency creates several risks. Even within a single vendor and a single hardware family, published fleet-level error summaries vary substantially across backends. From Table \ref{tab:ibm_heron_error_summary}, we observe a $\sim2.1\times$, a $\sim2.8$, and a $\sim6.9\times$ spread in the median 2Q error, layered 2Q error, and the median readout error, respectively, on the IBM Heron (133 qubit) backends. Since users typically do not control the exact device and routing used at execution time, these variations might prompt a cloud provider to divert jobs to error-prone qubits to conserve high-quality resources for premium workloads. Multi-tenant execution environments can expose users to unintentional interference, cross-talk, or even malicious co-tenants who intentionally introduce faults \cite{ghosh2023primer}. Compiler and allocator bugs may further degrade performance, yet the user has no mechanism to validate these decisions. As quantum computing grows, so does the need for forensic tools that reveal how a backend behaved during transpilation and execution, even when the provider does not disclose full internal information.

In this work, we take a step toward such forensic capability. We study whether backend-level error patterns can be reconstructed from the artifacts visible to end users, specifically, transpiled circuits and their associated structural features. Transpilation acts as an indirect probe of the hardware. The frequency with which certain qubits or links appear, the distribution of routing choices, and the sensitivity of a circuit to device layout all encode information about underlying error rates. Our goal is to extract this information with a data-driven model.

\subsection{Proposed Idea}
\label{subsec:proposed_idea}

We propose a forensic framework that analyzes the static topology of a backend together with dynamic features derived from many randomly generated and transpiled circuits. These features are processed using GNNs, which are well-suited to represent qubits and couplings as nodes and edges \cite{schuld2021machine,saravanan2022data}. The model learns how similar devices behave and, from that knowledge, estimates the per-qubit and per-link error rates of a new backend. Importantly, this method does not require calibration access to the target device. A user may run a collection of circuits, retrieve their transpiled versions, and apply our forensic model to assess the quality of the hardware that was implicitly selected for execution.

\subsection{Usage Scenario}
\label{subsec:usage}

\begin{figure}
    \includegraphics[width=0.5\textwidth]{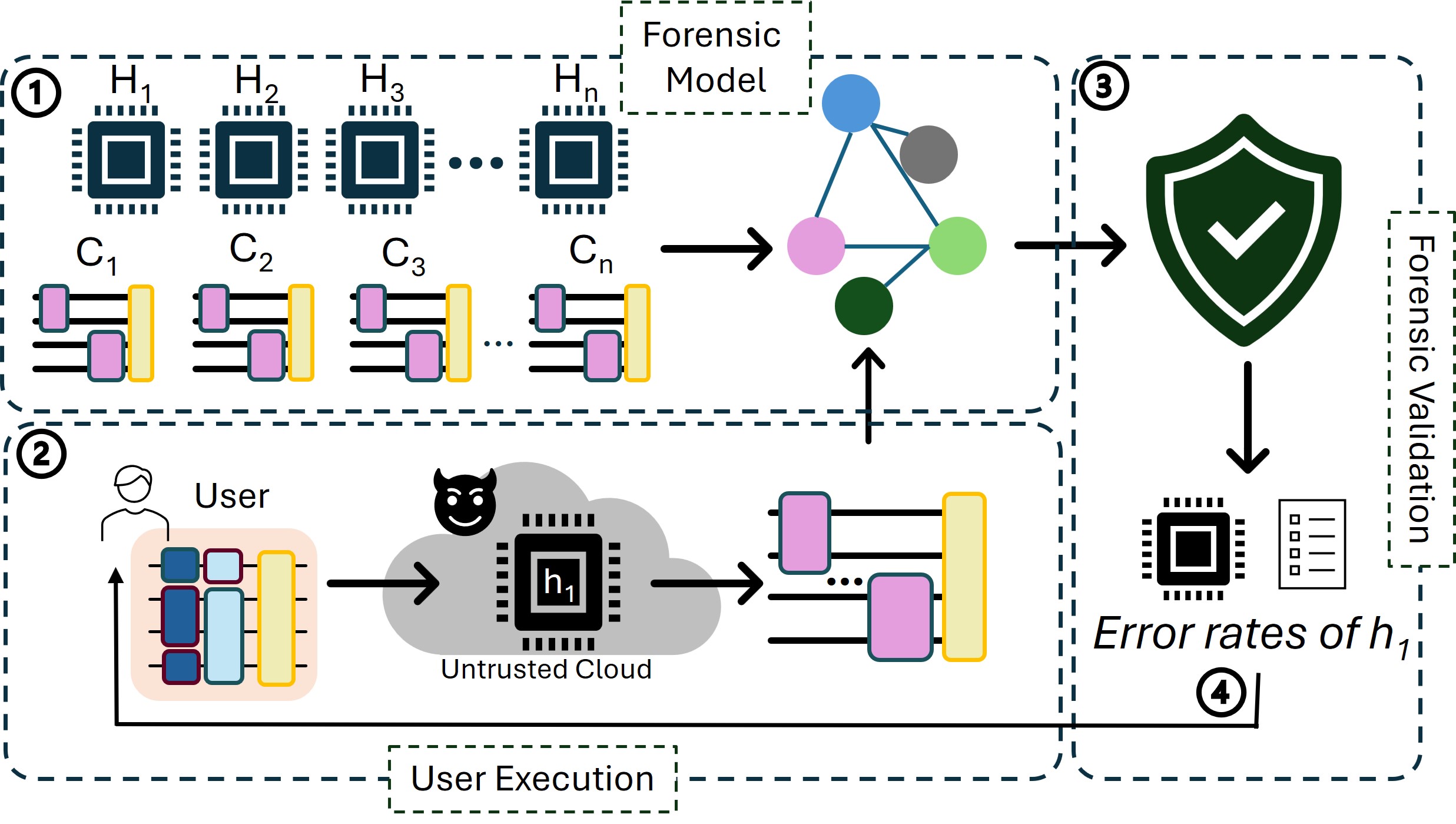}
    \centering
    \caption{A diagrammatic representation of the usage scenario of our framework. Part (1) shows the training of the forensic model. The GNN model trains on the error rates of qubits and qubit links from hardware $H_1$, $H_2$,...,$H_n$, and circuit pools $C_1$, $C_2$,...,$C_n$; (2) shows the user perspective where the user sends a quantum workload to an untrusted cloud and obtains a transpiled version. In (3), our proposed framework infers the error rates of the claimed execution backend $h_1$ from the transpiled circuits and its topology using the trained GNN to validate the provider’s claim. Finally, the predicted error rates are obtained and sent to the user in part (4). If the predicted values match the error rates claimed by the cloud provider, the claim is true, otherwise it is false.}
    \label{fig:flow}
\end{figure}

A realistic deployment of our forensic framework, as described in Fig. \ref{fig:flow}, involves two phases: an offline learning phase performed by a trusted analyst, and an online usage phase performed by an end user after running jobs on a cloud backend.

\textbf{Offline model preparation: }
A forensic analyst collects data from several publicly available or historical devices. For each backend, the analyst generates random circuits, transpiles them under fixed compiler settings, and extracts static and dynamic graph features. Since calibration data for these training backends are known, the analyst trains the GNN models to map graph features to one-qubit and two-qubit error rates. This produces a set of device-agnostic models that learn how compilation behavior reflects underlying hardware noise characteristics.

\textbf{User interaction with an untrusted cloud provider: }
A user submits quantum programs to a third-party cloud service and retrieves the transpiled circuits returned by the provider. Although the transpiled circuit reveals which physical qubits and couplings were used, the user has no visibility into why this mapping was chosen, whether other higher-quality mappings were available, or whether the device behaved according to the published calibration metrics. No calibration access or diagnostic measurements are required from the user's side.

\textbf{Forensic analysis: }
The user forwards the transpiled circuits (or the extracted features) along with the publicly documented topology of the target backend to the forensic tool. The trained models process these features and reconstruct an estimate of the backend’s node and edge error rates. Because the model was trained to generalize across backends, it does not rely on calibration information from the target device.

\textbf{Outcome: }
The forensic engine returns a predicted error map that highlights which qubits and couplings were likely strong, and which may have been weak or unreliable at the time of execution. Users can compare these reconstructed error patterns with the provider’s published specifications to detect inconsistencies, identify suspicious routing decisions, and evaluate whether their jobs were executed on hardware of the expected quality. This enables post-execution auditing in settings where transparency is limited and trust cannot be assumed.

\subsection{Applications of the Proposed Forensic Framework}

We outline several practical applications of the proposed forensic framework for different stakeholders in cloud quantum computing.

\textbf{User-side trust and auditing.}  
Users can apply the framework to assess whether their circuits were executed on hardware with the expected error characteristics. This is useful when observed performance deviates from the error information reported by the cloud provider. The analysis relies only on transpiled circuits and backend topology.

\textbf{Validation of calibration services.}  
Quantum cloud service providers often rely on third-party services to perform device calibration. The proposed framework can be used to verify whether the reported calibration data accurately reflects the underlying hardware behavior. This enables providers to detect calibration drift, misconfiguration, or incomplete calibration updates.

\textbf{Competitive analysis of new or undisclosed hardware.}  
When a new vendor or hardware platform enters the market, detailed error rates may not be publicly disclosed. Peer companies can use the framework to infer error maps from compilation behavior and topology information. This allows comparative analysis of hardware quality without access to proprietary calibration data.


\subsection{Contributions}\label{subsec:our_contributions}

\begin{enumerate}
    \item We motivate the need for data-driven quantum hardware forensics in cloud environments where allocation decisions are opaque and may expose users to reduced fidelity or adversarial behavior.
    \item We build a dataset across multiple IBM “Fake” 27-qubit backends and extract both static and transpilation-induced dynamic graph features that reflect hardware quality trends.
    \item We develop separate GNN-based regressors for one-qubit and two-qubit error estimation, designed to operate without calibration access to the target backend.
    \item We present a complete forensic workflow that users can apply after retrieving transpiled circuits, enabling independent reconstruction and auditing of backend error maps.
    \item We present a comparative analysis against prior forensic approaches and evaluate the robustness of our framework under realistic temporal noise drift.
\end{enumerate}

\subsection{Paper Organization}
\label{subsec:paper_organization}

Section II reviews background on quantum device architectures, transpilation, noise mechanisms, and GNNs. Section III presents the threat model, problem formulation, feature extraction process, and GNN architecture used for forensic inference. Section IV describes the experimental setup and reports experimental results, including error reconstruction accuracy, ablation studies, comparison with prior forensic work, and robustness under temporal noise variation. Section V concludes the paper and outlines future research directions.

\section{Background and Related Work} 
\label{sec:background}

\subsection{Quantum Device Architecture}

Superconducting quantum processors implement qubits arranged in a fixed interaction graph
$G = (V, E)$, where $V$ represents physical qubits and $(u,v) \in E$ denotes a hardware-supported two-qubit operation. IBM devices frequently adopt heavy-hex and related lattice geometries~\cite{IBM_heavyhex}, designed to reduce frequency collisions and crosstalk. Connectivity constraints have direct consequences: if a logical operation requires interaction between qubits that are not adjacent, the compiler must introduce SWAP gates, increasing circuit depth and cumulative noise.

Each qubit exhibits hardware-specific characteristics influenced by fabrication, control electronics, and spatial placement. Physical qubits with many direct neighbors in the coupling graph often become routing hubs, receiving more operations during compilation, while peripheral qubits experience distinct electromagnetic environments. These architectural factors shape the effective fidelity and execution behavior of compiled circuits.

\subsection{Quantum Error Mechanisms}

Quantum operations deviate from ideal unitaries due to decoherence, control inaccuracies, and environmental coupling. A noisy gate can be modeled as
\[
\mathcal{E}(\rho) = (1 - p)\, U\rho U^\dagger + p\, \eta(\rho),
\]
where $U$ is the intended unitary, $p$ is the gate error probability, and $\eta$ is a noise channel~\cite{Gardiner}.  
Single-qubit errors arise from amplitude damping, phase damping, and drive miscalibration. Two-qubit errors originate from imperfect entangling pulses, residual coupling, and crosstalk, and thus tend to dominate circuit failure rates.

Cloud providers publish periodic calibration metrics~\cite{IBMCAL} such as:  
\begin{itemize}
    \item \textbf{One-qubit gate error} $p_{1q}(i)$  
    \item \textbf{Two-qubit gate error} $p_{2q}(u,v)$  
    \item \textbf{Coherence indicators} ($T_1$, $T_2$)  
    \item \textbf{Readout error} $p_{\mathrm{ro}}(i)$  
\end{itemize}

These values drift over hours or days due to temperature changes, flux noise, and recalibration. Even devices with identical layouts exhibit heterogeneous error profiles, shaped by fabrication irregularities and control-line tuning.

\subsection{Quantum Circuit Transpilation}

A logical circuit must be transformed into an executable program that obeys both connectivity and native gate constraints. Modern transpilers such as Qiskit~\cite{Qiskit} perform:

\begin{enumerate}
    \item \textbf{Initial mapping} of logical qubits to physical qubits.
    \item \textbf{Routing} to insert SWAP operations to satisfy connectivity constraints.
    \item \textbf{Gate decomposition} into basis gate operations supported by the hardware.
    \item \textbf{Optimization} to reduce circuit depth, cancel adjacent gate operations, and improve fidelity.
\end{enumerate}

Two-qubit routing is particularly influential. The effective failure probability of a routed interaction across a path $\pi$ of couplings can be approximated by
\[
p_{\mathrm{eff}}(\pi) \approx 1 - \prod_{(u,v)\in\pi} (1 - p_{2q}(u,v)).
\]
Noise-aware routing~\cite{MuraliNoiseAware} may therefore avoid high-error couplings even if this increases path length. The resulting transpiled circuits encode observable patterns: preferred qubits, frequently used couplings, and regions systematically avoided due to noise.





\subsection{Graph Neural Networks}

Graph neural networks (GNNs) extend deep learning to domains where relationships between elements are encoded as edges. A GNN maintains node embeddings and updates them through message passing, in which each node aggregates information from its neighbors. After $k$ layers, the embedding of a node incorporates information from its $k$-hop neighborhood.

Common GNN architectures include:
\begin{itemize}
    \item \textbf{Graph Convolutional Networks (GCNs)} using normalized graph Laplacians~\cite{jiang2019semi},
    \item \textbf{GraphSAGE} with learnable aggregation functions and inductive generalization~\cite{GraphSAGE},
    \item \textbf{Graph Attention Networks (GATs)} employing attention weights to highlight informative neighbors~\cite{GAT}.
\end{itemize}

GNNs are widely used for node regression, link prediction, and physical-system modeling, where structural context influences local predictions~\cite{WuGNNsurvey}. Quantum processors map naturally to such graphs: qubits form nodes, couplings form edges, and both static properties (calibration) and dynamic properties (usage patterns) can be encoded as features for learning.

\subsection{Related Work}

Characterizing noise in quantum hardware traditionally relies on direct measurement techniques such as randomized benchmarking, gate set tomography, and cross-entropy benchmarking~\cite{RB, GST, CEB}. These methods provide accurate gate-level error estimates but require submitting large sets of diagnostic experiments to the target device. Such procedures assume full measurement access, stable calibration conditions, and dedicated runtime, which cloud users typically do not possess. Moreover, none of these methods can retrospectively assess the hardware state during a past execution.

Several efforts have examined indirect evidence of device behavior through transpiled circuits. Prior studies show that routing choices, SWAP insertion patterns, and coupling usage frequencies correlate with underlying hardware quality~\cite{rupshali_forensics_transpile}. These works demonstrate that compiler decisions reveal information about backend noise, and that transpiled circuits can be used to infer which regions of the device are stronger or weaker. Related forensic approaches have attempted to detect malicious or anomalous allocation policies or to infer partial device characteristics, but these methods do not reconstruct full per-qubit and per-edge error maps.

Machine learning has been applied to quantum hardware modeling and noise-aware compilation. Classical ML and GNN-based models have been used to predict circuit fidelity, guide qubit layout choices, optimize routing, and model backend performance~\cite{saravanan2022data}\cite{NoiseAdaptive_RL_GNN}. These approaches leverage device topology and calibration data to improve compilation or to approximate noise channels. However, they assume access to accurate calibration data for the target backend and focus on forward prediction (from hardware to performance) rather than forensic inversion.

Taken together, prior work shows that (i) compiler behavior carries meaningful information about hardware quality, and (ii) GNNs effectively learn structure-dependent noise patterns. However, no existing framework performs post-execution forensics by reconstructing the full node and edge error landscape of an unseen backend using only its topology and transpiled circuits. No prior method trains across multiple devices to generalize error inference to unknown hardware or treats transpilation statistics as a primary forensic signal. This gap motivates the approach developed in this work.

\section{Methodology}
\label{sec:Methodology}

\subsection{Threat Model and Trust Assumptions}

We study a setting where users execute quantum programs on cloud-based 
quantum hardware. These platforms offer access to several devices with 
various qubit technologies, coupling maps, and calibration properties, but 
their internal allocation and routing decisions are not fully visible. The user 
submits a circuit and receives only the transpiled version and execution results.

We consider two gaps in visibility. First, the user does not know the exact 
hardware conditions at the time of execution. Error rates provided by the 
vendor come from periodic calibration cycles, and they may not match the 
state of the backend when the job is routed. Second, the user cannot inspect 
the internal decisions made by the mapper or optimizer. The transpiler may 
select a different subgraph, route through unexpected couplings, or avoid 
high-noise qubits in ways that are not explicitly disclosed.

Our threat model allows both benign and adversarial deviations from expected 
behavior. A cloud provider may reassign jobs for load balancing, reduce 
resource usage by favoring less reliable qubits, or apply proprietary routing 
policies that the user cannot audit. An insider adversary or a compromised 
allocation policy may intentionally divert circuits to higher-error qubits to 
degrade fidelity or reduce operational cost. The user, however, still receives 
the transpiled circuit, which preserves the chosen physical qubits, the set of 
active couplings, and the gate counts along each edge.

Our forensic model operates under realistic constraints that match the data 
accessible in practice. We assume:
\begin{itemize}
    \item The user does \emph{not} have access to direct calibration data of the backend under investigation. This is justified under two situations, (i) the user cannot fully trust the hardware where the program is likely executed (which is the proposed threat model) and (ii) even though the hardware matches the user expectation, the coupling map may not be exactly same as expected.
    \item The user \emph{does} know the hardware topology (i.e., the qubit 
    connectivity graph), which is publicly documented. This is justified since changing the hardware may result into significant changes in execution profile that can be easily detected. User may also employ previously proposed approaches ~\cite{rupshali_forensics_transpile} to detect the hardware. 
    \item The user can retrieve the transpiled quantum circuit produced by the 
    provider. This is in accordance with the current practice followed by every publicly available quantum service providers.
    \item A forensic service maintains a collection of backends for which 
    both calibration data and transpiled circuit statistics are available. These 
    devices form the training corpus. This is a justified assumption since forensic service provider may access historical data available in literature and/or vendor website.
\end{itemize}

The key assumption is that hardware quality shapes transpilation outcomes. 
Noise-aware compilers route around weak couplings, place heavily interacting 
logical qubits on high-fidelity physical nodes, and bias CX usage toward 
reliable links. These patterns are visible in aggregated gate statistics extracted 
from transpiled circuits. Our goal is to invert this relationship. Given only the 
topology and the transpiled circuit features of an unseen backend, we estimate 
its per-qubit and per-edge error rates, enabling post-execution forensic 
auditing without calibration access.

\subsection{System Overview}
Our framework reconstructs the error map of an unseen quantum backend using only public topology and workload-dependent artifacts obtainable from transpilation. The pipeline consists of an offline learning stage and an online inference stage. In training, we assemble a set of backends with available calibration-derived error rates and generate multiple circuit pools per backend. Each pool is transpiled under fixed compiler settings, and we extract graph-structured features comprising (i) static topology descriptors and (ii) dynamic usage statistics aggregated over the transpiled pool. The resulting samples pair a backend graph with node and edge labels corresponding to single-qubit and two-qubit error metrics. We train two GNN regressors, one specializing in per-qubit error prediction and the other in per-coupling error prediction, using robust regression objectives and mask-aware evaluation to handle missing calibration entries. At inference time on a target backend, the user-visible transpiled circuits are processed by the same feature pipeline to form graph inputs without requiring calibration access. The trained models produce node- and edge-level error estimates, which are aggregated across multiple pools to reduce workload-specific variance; a lightweight validation-fitted linear correction is optionally applied to mitigate cross-device scale/offset bias. The output is a reconstructed error map over qubits and couplings, enabling backend auditing and quality assessment from transpilation behavior alone. The technical details are discussed in the following sections.

\subsection{Problem Formulation}
\label{sec:problem_formulation}

We model a quantum backend $b$ as an attributed graph
\[
G^{(b)} = \big(V^{(b)}, E^{(b)}\big),
\]
where $V^{(b)}=\{1,\dots,N_b\}$ denotes physical qubits and $E^{(b)} \subseteq V^{(b)} \times V^{(b)}$ denotes the set of hardware-supported two-qubit couplings.
Each backend has an (unknown) \emph{error map} consisting of per-qubit and per-coupling error rates:
\[
\mathbf{y}_V^{(b)} \in \mathbb{R}^{N_b}, 
\qquad 
\mathbf{y}_E^{(b)} \in \mathbb{R}^{|E^{(b)}|},
\]
where $y_V^{(b)}(v)$ represents a single-qubit error metric (e.g., average 1q gate error) for qubit $v$, and $y_E^{(b)}(u,v)$ represents a two-qubit error metric (e.g., CX error) for coupling $(u,v)$.

At inference time on a target backend $b^\star$, the user is assumed to know the topology $G^{(b^\star)}$ (public connectivity) and can retrieve transpiled circuits, but does \emph{not} have trusted access to the target error map $\big(\mathbf{y}_V^{(b^\star)},\mathbf{y}_E^{(b^\star)}\big)$.
During offline training, a forensic analyst has access to a set of training backends $\mathcal{B}_{\text{train}}$, for which calibration-derived labels are available.

\textbf{Circuit pools and transpilation.}
For each backend $b$, we generate multiple circuit pools
\[
\mathcal{C}_{p}^{(b)} = \{ C_{p,1},\dots,C_{p,M}\},
\]
and transpile them under fixed transpiler settings (routing policy, optimization level, seed, and basis constraints).
The transpiled pool $\widetilde{\mathcal{C}}_{p}^{(b)}$ constitutes the user-visible artifact from which we extract dynamic usage statistics.

\textbf{Feature construction.}
We build node and edge feature matrices by combining (i) static backend descriptors and (ii) dynamic transpilation-induced statistics:
\begin{align*}
\mathbf{X}_{V,p}^{(b)} &= 
\big[ \mathbf{X}_{V}^{(b,\text{stat})} \,\|\, \mathbf{X}_{V,p}^{(b,\text{dyn})}\big]
\in \mathbb{R}^{N_b \times F_V},\\
\mathbf{X}_{E,p}^{(b)} &= 
\big[ \mathbf{X}_{E}^{(b,\text{stat})} \,\|\, \mathbf{X}_{E,p}^{(b,\text{dyn})}\big]
\in \mathbb{R}^{|E^{(b)}| \times F_E}.
\end{align*}
Static features $\mathbf{X}_{V}^{(b,\text{stat})}, \mathbf{X}_{E}^{(b,\text{stat})}$ depend only on the backend graph and published properties (e.g., degree/position indicators and optional calibration-like priors, when available).
Dynamic features $\mathbf{X}_{V,p}^{(b,\text{dyn})}, \mathbf{X}_{E,p}^{(b,\text{dyn})}$ are empirical statistics computed from $\widetilde{\mathcal{C}}_{p}^{(b)}$; i.e., each node/edge feature is an aggregate (e.g., mean/ratio/count) over transpiled circuits in the pool.

\textbf{Handling heterogeneous backends.}
To support backends with differing qubit counts and coupling sets, we include binary masks
\begin{equation*}
\mathbf{m}_V^{(b)} \in \{0,1\}^{N_b}, \qquad \mathbf{m}_E^{(b)} \in \{0,1\}^{|E^{(b)}|},
\end{equation*}
indicating which nodes/edges exist and should contribute to training loss and evaluation.

\textbf{Learning objective (graph-to-error-map regression).}
Each pool yields a supervised training sample
\begin{equation*}
\mathcal{S}_{p}^{(b)} =
\Big(G^{(b)},\, \mathbf{X}_{V,p}^{(b)},\, \mathbf{X}_{E,p}^{(b)},\, \mathbf{m}_V^{(b)},\, \mathbf{m}_E^{(b)},\, \mathbf{y}_V^{(b)},\, \mathbf{y}_E^{(b)}\Big),
\end{equation*}
for $b \in \mathcal{B}_{\text{train}}$.
We seek inductive predictors (implemented as GNNs) that output node- and edge-level estimates:
\begin{align*}
\widehat{\mathbf{y}}_V^{(b)} &= f_V\!\left(G^{(b)}, \mathbf{X}_{V,p}^{(b)}, \mathbf{X}_{E,p}^{(b)}\right),\\
\widehat{\mathbf{y}}_E^{(b)} &= f_E\!\left(G^{(b)}, \mathbf{X}_{V,p}^{(b)}, \mathbf{X}_{E,p}^{(b)}\right),
\end{align*}
trained on $\mathcal{B}_{\text{train}}$ and evaluated on a fully held-out backend $b^\star$ where labels are never used during training.
At inference time, the same pipeline forms $\mathbf{X}_{V,p}^{(b^\star)}$ and $\mathbf{X}_{E,p}^{(b^\star)}$ from topology and transpiled pools, producing a reconstructed error map $\big(\widehat{\mathbf{y}}_V^{(b^\star)}, \widehat{\mathbf{y}}_E^{(b^\star)}\big)$.

\subsubsection{Static Graph Features}
\label{sec:static_features}

Static features are computed once per backend from the undirected coupling graph
$G^{(b)}$ and are independent of any circuit pool.

\textbf{Node static features: }
For each node $v\in V^{(b)}$, we construct
$\mathbf{X}_{V,p}^{(b,\text{stat})}$ as a concatenation of standard topology descriptors: (i) the degree, normalized as $x_{\deg}(v) = \frac{\deg(v)}{\max_{w\in V^{(b)}} \deg(w)}$; (ii) the Betweenness Centrality and the Clustering Coefficient; (iii) the harmonic centrality, $x_{\mathrm{harm}}(v) = \sum_{w\in V^{(b)}\setminus\{v\}} \frac{1}{d(v,w)},$, with $d(v,w)$ being the shortest-path distance between qubits $v$ and $w$; and (iv) the normalized value of $k-$core number, $x_{\mathrm{core}}(v) = \frac{\kappa(v)}{\max_{w\in V^{(b)}} \kappa(w)}$. The degree and $k-$core values capture the immediate connectivity and the reflection of whether a qubit lies in the densely connected portion of the backend, respectively. Centrality features are considered as a measure of the topological position of a qubit with respect to the routing paths. 


\textbf{Edge static features: }
For each coupling $e=(u,v)\in E^{(b)}$, we fix a canonical undirected edge ordering
$\{e_i=(u_i,v_i)\}_{i=1}^{|E^{(b)}|}$ with $(u_i,v_i)$ stored as $(\min(u_i,v_i),\max(u_i,v_i))$, to ensure consistent alignment between the static and dynamic edge features. We construct
$\mathbf{X}^{(b,\mathrm{stat})}_{E,p}$ as a concatenation of:
(i) the Edge Betweenness Centrality of an edge, (ii) degree-based endpoint summaries, $x_{\Sigma\deg}(e) = \mathrm{norm}\!\big(\deg(u)+\deg(v)\big)$ and $x_{\Pi\deg}(e) = \mathrm{norm}\!\big(\deg(u)\deg(v)\big)$ and (iii) a binary bridge indicator, $x_{\mathrm{bridge}}(e) = \mathbb{I}\{e\ \text{is a bridge}\}$
where $\mathrm{norm}(a)=(a-\min a)/(\max a-\min a)$ is min--max normalization applied over edges within the backend graph. The centrality and degree-based summaries capture the positional feature of an edge, and the binary bridge feature indicates if removing the edge increases the number of connected components.

\subsubsection{Dynamic Graph Features}
\label{sec:dynamic_features}

Dynamic features summarize how a fixed transpilation policy exercises each qubit and coupling under a given workload pool.
For a pool $\widetilde{\mathcal{C}}_{p}^{(b)}$, we extract per-circuit counts by scanning the transpiled gate list and counting (a) single-qubit operations and (b) supported two-qubit operations (e.g., CX/ECR/SWAP-like gates) on physical qubits/couplings.

\textbf{Node dynamic features: }
For each circuit $\widetilde{C}_{p,i}$ we compute the number of 1-qubit gates, $g^{(i)}_{1}(v)$, and a binary indicator of a node participating in a 1-qubit or 2-qubit gate, $u^{(i)}(v) = \mathbb{I}$. From these primitives, we construct $\mathbf{X}^{(b,\mathrm{dyn})}_{V,p}\in\mathbb{R}^{N_b\times 3}$, having three columns, representing: (i) the node coverage using $u^{(i)}(v)$; (ii) the share of 1-qubit workloads, $\frac{g^{(i)}_{1}(v)}{\max(1,G^{(i)}_{1})}$, $G^{(i)}_1$ being the sum of $g_1^{(i)}(v)$ over all nodes; and (iii) the relative 1-qubit workload share, $\frac{g^{(i)}_{1}(v)}{\max(1,\mu^{(i)}_{1})}$, where $\mu^{(i)}_{1} := \frac{1}{|A^{(i)}|}\sum_{v\in A^{(i)}} g^{(i)}_{1}(v),
A^{(i)} := \{v\in V^{(b)}: u^{(i)}(v)=1\}$.

\textbf{Edge dynamic features: }Similar to the node dynamic features, we compute the number of 2-qubit gates acting on an edge $e$ as $g^{(i)}_{2}(e)$, and a binary indicator for each edge, $v^{(i)}(e) = \mathbb{I}\{g^{(i)}_{2}(e) > 0\}$. From these primitives, we construct $\mathbf{X}^{(b,\mathrm{dyn})}_{E,p}\in\mathbb{R}^{|E^{(b)}|\times 2}$, having two columns, representing: (i) the edge coverage using $v^{(i)}(e)$; and (ii) the share of 2-qubit workloads $\frac{g^{(i)}_{2}(e)}{\max(1,G^{(i)}_{2})}$, where $G^{(i)}_{2}$ is the sum of $g^{(i)}_{2}(e)$ over all edges. 

These features are computed per circuit and normalized (e.g., by total gate counts or active-qubit averages) before being averaged over the pool, making them comparable across circuits with different sizes and depths.


\subsection{Dataset Construction}


For each 27-qubit backend $b$, we construct an evaluation dataset by generating $P$ circuit pools, each containing $M$ randomly synthesized quantum circuits designed to induce nontrivial placement and routing effects under transpilation. One such pool is a supervised graph sample (same topology and labels, but different dynamic statistics). Each circuit is created by sampling an active width $w\in\{1,\dots,N_b\}$, $N_b=27$ and a target two-qubit budget uniformly sampled from the unidirected edge ordering discussed earlier. The circuit is composed using an alternate structure of 1-qubit rotation layers, and 2-qubit CNOT gate on a random matching of the active set of qubits (having a 2-qubit link) till the depth cap is reached. Each datapoint is generated using a deterministic seed, making the dataset regenerable. Transpilation is done using the Qiskit 1.2.2 transpiler with the optimization level set to 1. The node and edge features (discussed in Section III.C) are then extracted using the transpiled circuits, followed by per-circuit normalization and pool averaging.
Supervision targets are obtained from the backend properties: node labels $y_V^{(b)}(v)$ are defined as the mean reported gate errors over all single-qubit gates acting on physical qubit $v$, and edge labels $y_E^{(b)}(u,v)$ are defined as the mean reported gate error over all two-qubit gates whose operand pair corresponds to coupling $(u,v)$ (after canonical undirected edge normalization); missing calibration entries are treated as undefined and excluded via masks during training and evaluation. Finally, feature standardization (z-scoring) is fit using only pools from $\mathcal{B}_{\mathrm{train}}$ and applied unchanged to the held-out backend $b^\star$, yielding an inductive, device-level generalization test where the GNN reconstructs $\big(\mathbf{y}_V^{(b^\star)},\mathbf{y}_E^{(b^\star)}\big)$ from the topology and transpiled artifacts alone.

\subsection{GNN Architecture}
\label{sec:gnn_training}
As discussed before, given a backend graph and pool-specific node and edge features, our goal is to regress the node- and edge-level error maps $\mathbf{y}^{(b)}_V$ and $\mathbf{y}^{(b)}_E$ for an unseen backend using only topology and transpilation-derived usage statistics.

We implement a GraphSAGE-type, edge-aware message passing regressor that first embeds raw features using shallow MLP encoders and then propagates information over the coupling graph. Specifically, for each qubit $v\in V^{(b)}$ and coupling $e=(u,v)\in E^{(b)}$, we compute initial latent representations
\begin{equation*}
\mathbf{h}^{(0)}_{v}=\phi_V(\mathbf{X}_v)\in\mathbb{R}^{H},\qquad
\mathbf{e}_{e}=\phi_E(\mathbf{X}_{e})\in\mathbb{R}^{H},
\end{equation*}
where $\mathbf{X}_v$ is the concatenated static+dynamic node feature vector for $v$ and $\mathbf{X}_{e}$ is the concatenated static and dynamic edge feature vector for $(u,v)$. We then form edge-conditioned messages using both endpoint embeddings and the edge embedding,
\begin{equation*}
\mathbf{m}_{uv}=\phi_M\big([\mathbf{h}^{(0)}_{u}\,\|\,\mathbf{h}^{(0)}_{v}\,\|\,\mathbf{e}_{e}]\big)\in\mathbb{R}^{H},
\end{equation*}
and aggregate them symmetrically at each endpoint using degree-normalized mean aggregation:
\(
\mathbf{a}_{v}=\frac{1}{\max(1,\deg(v))}\sum_{(v,w)\in E^{(b)}} \mathbf{m}_{vw}.
\)
The node embedding is updated by combining the local representation with the aggregated neighborhood signal,
\(
\mathbf{h}^{(1)}_{v}=\phi_U\big([\mathbf{h}^{(0)}_{v}\,\|\,\mathbf{a}_{v}]\big),
\)
where $\phi_V,\phi_E,\phi_M,$ and $\phi_U$ are MLPs with nonlinearities and dropout. From these embeddings, we predict node-level and edge-level error metrics using task-specific regression heads. The node head is linear in the final node embedding,
\[
\hat{z}_V(v)=g_V(\mathbf{h}^{(1)}_{v}),
\]
whereas the edge head combines the endpoint representations with the edge embedding through an MLP,
\[
\hat{z}_E(u,v)=g_E\big([\mathbf{h}^{(0)}_{u}\,\|\,\mathbf{h}^{(0)}_{v}\,\|\,\mathbf{e}_{e}]\big).
\]
We train two \emph{independent} regressors, one specialized for single-qubit errors and one specialized for coupling errors, since the two targets exhibit distinct numerical scales and distributions.

\textbf{Output Transforms: }
Single-qubit error rates are typically low, and direct regression in the original scale can under-resolve differences in the $10^{-4}$ regime. Accordingly, we train the node regressor in log-space and invert the transform during inference time.
For edge targets, we regress in the original scale while enforcing non-negativity through a softplus output,
which reflects the fact that two-qubit error rates are nonnegative and typically larger and more variable than one-qubit errors.

\textbf{Loss function: }
Both models are trained with a masked Huber loss, which is robust to outliers and provides stable gradients across heterogeneous backends. For the \emph{node} model we minimize
\begin{equation*}
\mathcal{L}_V=\frac{1}{\sum_{v} m_V^{(b)}(v)}\sum_{v\in V^{(b)}} m_V^{(b)}(v)\,
\rho_{\delta_V}\!\Big(\hat{z}_V(v)-\log(1+y_V(v))\Big),
\end{equation*}
where $z_V(v)-\log\big(1+y_V(v)\big)$ is the log-space representation
Similarly, for the \emph{edge} model we minimize
\begin{equation*}
\mathcal{L}_E=\frac{1}{\sum_{e} m_E^{(b)}(e)}\sum_{(u,v)\in E^{(b)}} m_E^{(b)}(u,v)\,
\rho_{\delta_E}\!\Big(\hat{y}_E(u,v)-y_E(u,v)\Big),
\end{equation*}
where $\hat{y}_E(u,v)$ represents the softplus of $\hat{z}_E(u,v)$, 
$\rho_{\delta}(\cdot)$ denotes the Huber penalty with threshold $\delta$, and the masks $m_V^{(b)}$ and $m_E^{(b)}$ exclude undefined labels (e.g., missing calibration entries) from contributing to the objective.
When transferring across devices, predictions can exhibit a consistent global scale and offset bias. To correct this without using any calibration information from the held-out backend, we fit a lightweight linear mapping on validation predictions from the training backends,
\(
y \approx a\,\hat{y} + b,
\)
where $(a,b)$ are obtained by least squares on the validation set (performed separately for node and edge models). At inference time on the hold-out backend $b^\star$, we apply the learned calibration to the raw predictions and clamp to nonnegative values, yielding the final reconstructed error maps $\widehat{\mathbf{y}}^{(b^\star)}_V$ and $\widehat{\mathbf{y}}^{(b^\star)}_E$.

\subsection{Holdout Inference and Error Map Reconstruction}
\label{sec:holdout_inference}

For the holdout backend $b^\star$, we generate $P^\star$ transpiled circuit pools under the same fixed compilation policy used during training and extract the corresponding feature tensors $\{\mathbf{X}^{(b^\star)}_{V,p},\mathbf{X}^{(b^\star)}_{E,p}\}_{p=1}^{P^\star}$. Each pool $p$ is treated as an independent graph sample on the common topology $G^{(b^\star)}$ and is processed by the trained node and edge regressors (Sec.~\ref{sec:gnn_training}) to produce pool-wise predictions $\widehat{\mathbf{y}}^{(b^\star)}_{V,p}\in\mathbb{R}^{N_{b^\star}}$ and $\widehat{\mathbf{y}}^{(b^\star)}_{E,p}\in\mathbb{R}^{|E^{(b^\star)}|}$. To obtain a backend-level forensic estimate that is less sensitive to workload-specific idiosyncrasies, we aggregate predictions across pools by simple averaging,
\begin{equation*}
\widehat{\mathbf{y}}^{(b^\star)}_{V}=\frac{1}{P^\star}\sum_{p=1}^{P^\star}\widehat{\mathbf{y}}^{(b^\star)}_{V,p},\qquad
\widehat{\mathbf{y}}^{(b^\star)}_{E}=\frac{1}{P^\star}\sum_{p=1}^{P^\star}\widehat{\mathbf{y}}^{(b^\star)}_{E,p}.
\end{equation*}
We then apply the validation-fitted linear calibration (learned exclusively from training-backend validation predictions) to correct systematic cross-device scale bias, i.e.,
\begin{equation*}
\widehat{\mathbf{y}}^{(b^\star)}_{V}\leftarrow \mathrm{clip}_+\!\big(a_V\,\widehat{\mathbf{y}}^{(b^\star)}_{V}+b_V\big),
\widehat{\mathbf{y}}^{(b^\star)}_{E}\leftarrow \mathrm{clip}_+\!\big(a_E\,\widehat{\mathbf{y}}^{(b^\star)}_{E}+b_E\big),
\end{equation*}
where $\mathrm{clip}_+(x)=\max\{0,x\}$ is applied elementwise. For evaluation purposes only, we compare these reconstructed error maps against the calibration-derived ground truth $\big(\mathbf{y}^{(b^\star)}_{V},\mathbf{y}^{(b^\star)}_{E}\big)$. The detailed results are presented in the following section.






\section{Experimental Evaluation}

Here we present the evaluation of our proposed framework.

\textbf{Benchmarking: }Since the forensic framework is applicable to any user workload irrespective of the algorithm, we demonstrate the working on randomly generated circuits. We consider a pool of 1000 such circuits that are executed on one 27-qubit IBM backend to be one sample. The training data is accumulated from 100 such pools across five backends. 
All the circuits are transpiled using identical Qiskit 1.2.2 settings (fixed random seed and coupling map, and optimization level=1). To effectively demonstrate the efficacy of our approach, we consider five such backends (viz., \emph{FakeAlgiers, FakeAuckland, FakeCairo, FakeKolkata, and FakeMontreal}), train on the first four backends, and designate the last one as the \emph{holdout device}, whose calibration values are used for final evaluation.

\textbf{GNN Hyperparameters: }Two GNNs are separately trained for node and edge prediction. The models are 3-layered, having a hidden dimension of 64, trained using the Adam optimizer, with a base learning rate of $10^{-3}$. We find that training the model for node prediction for 22 epochs and edge prediction for 33 epochs is optimal through early-stopping using the RMSE score.
We use only the validation data from training backends to fit the linear calibration model for two-qubit predictions. Convergence takes at most 4s at the longest. For the target backend, we generate the same number of circuit pools and extract dynamic features, and predictions are made for each pool, which is averaged to produce a single reconstructed error vector. 

\textbf{Experimental Setup: }All the algorithms for this work have been implemented in Python \textit{3.9.19} on an Intel Core \textit{Ultra 7 155U} CPU with a clock frequency of \textit{1.70} GHz.

We attempt to answer the following research questions through our evaluation:
\begin{itemize}
    \item \textbf{RQ1: }How accurately can our framework reconstruct per-qubit and per-coupling error rates compared to ground-truth calibration values?
    \item \textbf{RQ2: }Which knobs (e.g., static/dynamic features, volume of data) in the framework affect performance the most?
    \item \textbf{RQ3: }How does our framework compare against prior research?
    \item \textbf{RQ4: }How robust is the inference under temporal variation of noise?
\end{itemize}

\subsection{(RQ1) Reconstructing error rates}

\begin{figure}[t]
    \includegraphics[width=0.5\textwidth]{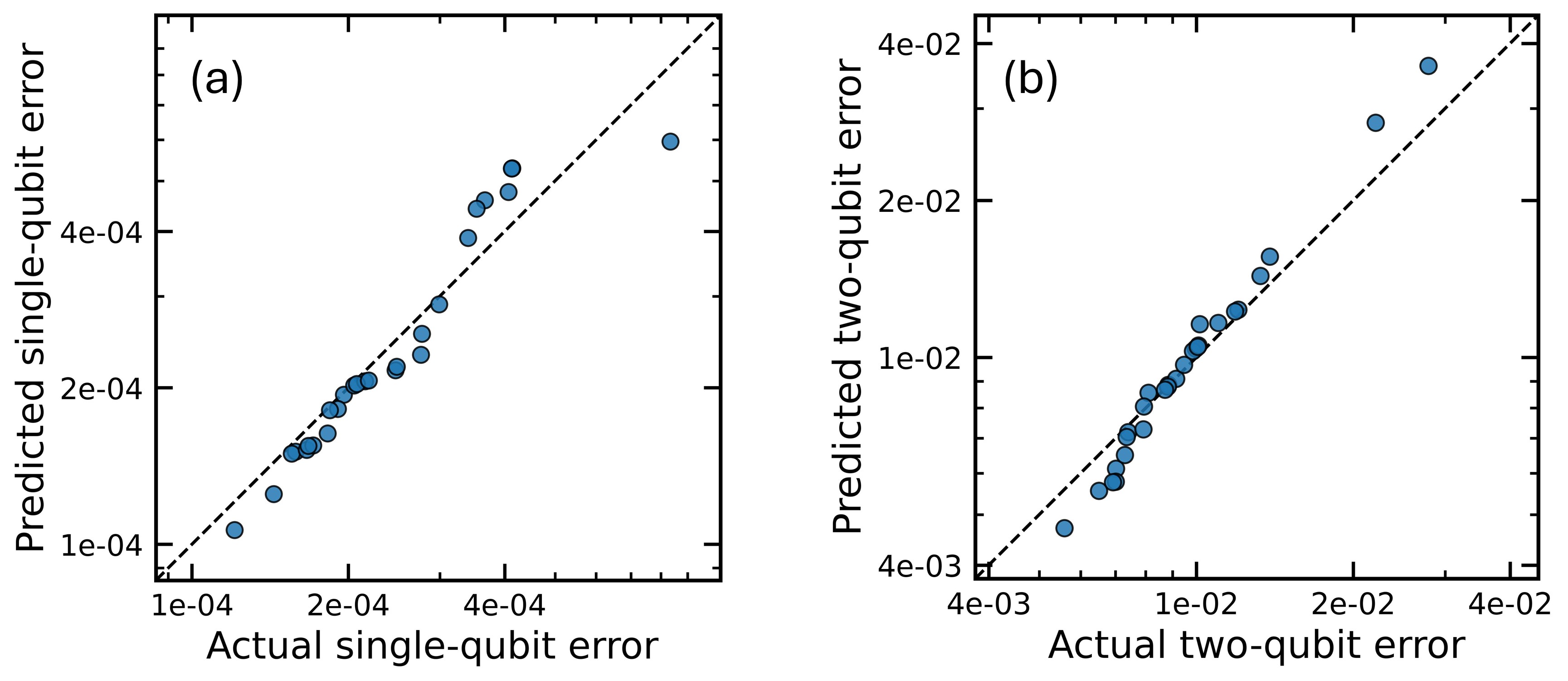}
    \centering
    \caption{Predicted vs. actual error rates for an unseen holdout backend. (a) Single-qubit error prediction at the node level. (b) Two-qubit error prediction at the edge level.}
    \label{fig:node_edge}
\end{figure}

We assess reconstruction of single-qubit and qubit-link error rates on the holdout backend by running inference strictly under the threat model constraints, using GNNs trained on data from the remaining backends. Fig. \ref{fig:node_edge}(a) and (b) compare predicted values against the holdout backend’s ground-truth calibration, with the identity line $\hat{y}=y$ indicating perfect reconstruction; points near the diagonal imply close agreement, while deviations reflect over- or under-estimation. Both axes are shown on a logarithmic scale to expand the low-error regime and avoid clustering near the origin. The results exhibit a clear monotonic alignment along the identity line, indicating that the model learns a consistent mapping from backend features to noise characteristics and generalizes well to an unseen device. Quantitatively, the average percent difference between predicted and actual values is approximately 22\% for single-qubit errors and 18\% for qubit-link errors on the holdout backend. This is quite good result considering the raw error rates are extremely small. A small number of outliers appear in the high-error tail, corresponding to the weakest qubits/couplings on the holdout backend, where reconstruction is inherently more difficult because noise-aware transpilation avoids these components and thus provides sparse, partially censored evidence that can saturate near zero usage. In this regime, multiple weak components may become difficult to distinguish due to similar feature signatures and additional nonlinearity introduced by global routing constraints and alternative-path availability, although the overall monotonic trend remains consistent and deviations are confined to a limited extreme subset.

\begin{table}[t]
\centering
\caption{Ranking consistency between actual and predicted error rates.}
\label{tab:ranking_consistency}
\begin{tabular}{lcc}
\hline
Component & Spearman $\rho$ & Top-10 Overlap \\
\hline
Nodes & 0.98 & 9 / 10 \\
Edges & 0.96 & 8 / 10 \\
\hline
\end{tabular}
\end{table}

We also evaluate how closely the predicted error rates match the ranking obtained from the actual error rates using two rank-based metrics. We first report Spearman’s rank correlation coefficient (\(\rho\)) \cite{ali2022spearman}, which measures agreement between the rankings defined by actual and predicted errors. A value of \(\rho = 1\) indicates identical ordering, while lower values indicate increasing mismatch. To focus on the most error-prone components, we also compute the Top-10 overlap, defined as the number of components that appear in the top ten highest-error ranks under both actual and predicted errors. This metric directly reflects whether the same high-risk nodes and edges are identified. As shown in Table~\ref{tab:ranking_consistency}, the predicted rankings closely match the actual rankings for both nodes and edges. Nodes achieve a Spearman correlation of 0.98 with a Top-10 overlap of 9 out of 10, while edges achieve a correlation of 0.96 with a Top-10 overlap of 8 out of 10. These results show that the model preserves both global ranking structure and the identification of the most error-prone components.

\begin{rqsummarybox}
\textbf{RQ1 Summary: }Our proposed framework reconstructs the holdout backend's per-qubit and per-coupling error rates with strong agreement to calibration (points track $\hat{y}=y$), achieving average percent differences of $\sim$22\% and $\sim$18\%, respectively.
\end{rqsummarybox}

\subsection{(RQ2) Ablation Study}

We isolate the impact of the volume of data and backend features on the performance of our framework. We use a scale-invariant \emph{log-ratio mismatch} metric to quantify the reconstruction quality of our framework while varying the number of pools of circuits available to the user or the number of backends available to the user. For each component $i$ (a physical qubit node or a coupling edge), let $\hat{y}_i$ denote the predicted error rate and $y_i$ the ground-truth calibration error rate. We first define the ratio error
\(
r_i = \frac{\hat{y}_i}{y_i}.
\)
Because both over- and under-estimation should be penalized symmetrically, we measure the magnitude of the log-ratio:
\[
m_i = \left|\log(r_i)\right|.
\]
We then average this mismatch across all nodes and edges, respectively:
\[
\mathcal{M}_{\mathrm{nodes}} = \frac{1}{N}\sum_{i \in \mathcal{V}} m_i, 
\qquad
\mathcal{M}_{\mathrm{edges}} = \frac{1}{E}\sum_{i \in \mathcal{E}} m_i,
\]
where $N=|\mathcal{V}|$ is the number of physical qubits and $E=|\mathcal{E}|$ is the number of couplings. Since the per-qubit error rates are of the order $10^{-4}$ and the qubit-link error rates are of the order $10^{-2}$, and the predicted values, $\hat{y}_i$ are \emph{not} a factor too large or too small of the original value $y_i$, a log-ratio mismatch metric is sufficient to capture the error landscape of our proposed framework. 

\begin{figure}[t]
    \includegraphics[width=0.5\textwidth]{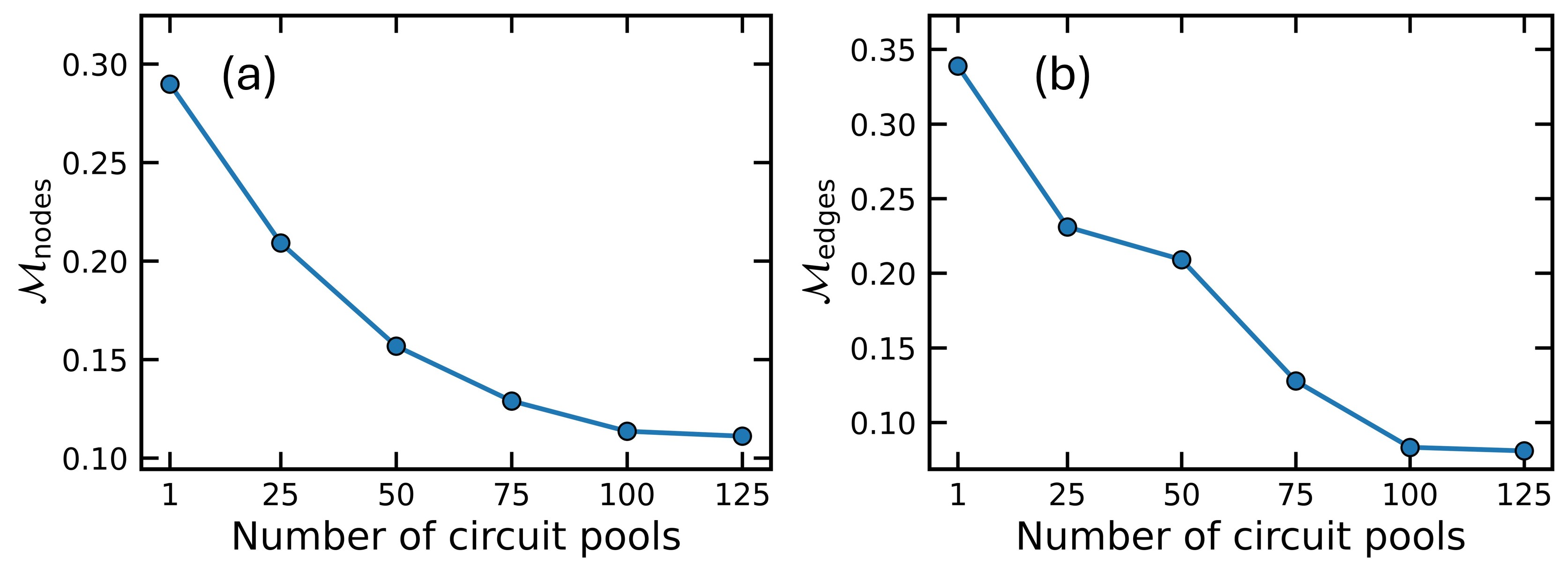}
    \centering
    \caption{Average log-ratio mismatch for (a) nodes and (b) edges as a function of the number of circuit pools used to extract dynamic features. Each pool contains 1000 circuits.}
    \label{fig:pool}
\end{figure}

\begin{figure}[t]
    \includegraphics[width=0.5\textwidth]{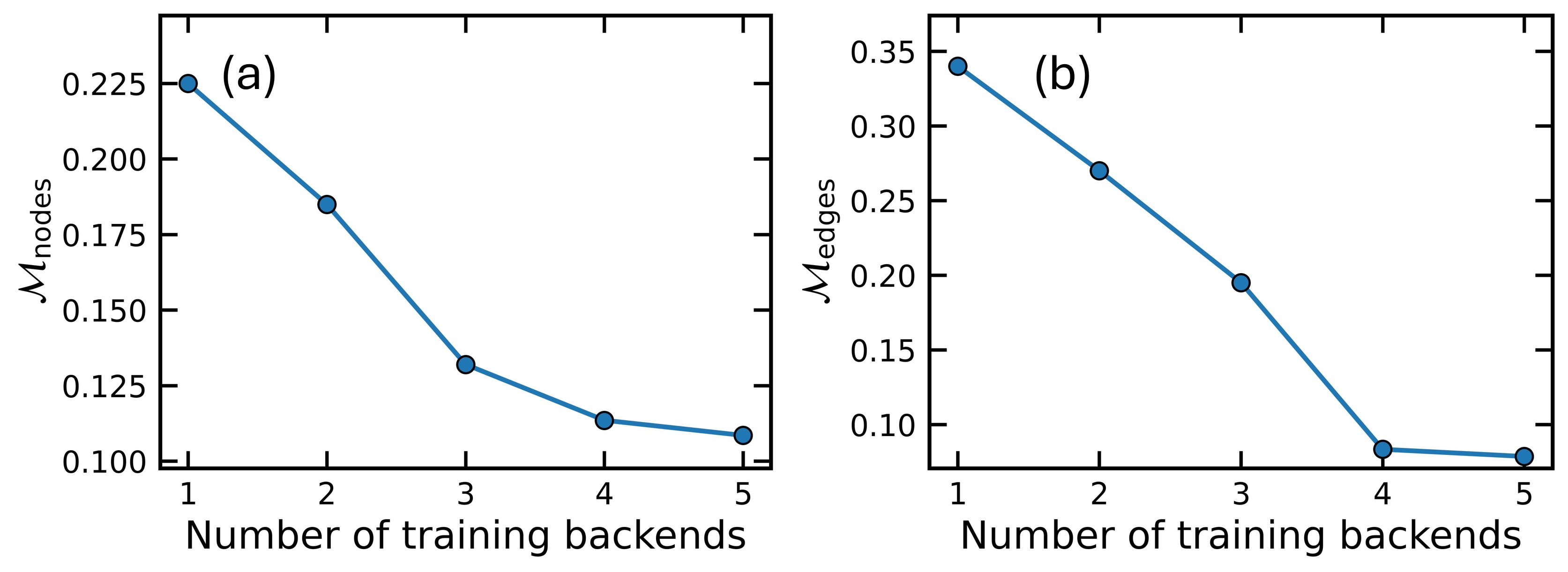}
    \centering
    \caption{Effect of the number of training backends on reconstruction accuracy. Average log-ratio mismatch for (a) nodes and (b) edges decreases as more backends are used during training.}
    \label{fig:backend}
\end{figure}

Fig. \ref{fig:pool} illustrates the effect of data volume on reconstruction quality. For both nodes and edges, the average log-ratio mismatch decreases as more circuit pools are used, indicating more accurate error prediction.
The largest improvement occurs when moving from a single pool to a small number of pools. This shows that dynamic feature estimates become more reliable once circuit statistics are aggregated across multiple pools. As the number of pools increases further, the mismatch continues to decrease but at a slower rate, indicating diminishing returns.
Edge mismatches are higher than node mismatches at low pool counts, reflecting the higher variability of two-qubit operations. With sufficient pools, both node and edge mismatches converge to low values, demonstrating that the framework benefits from increased circuit diversity. 
The implication of the above study is that the forensic service provider would require a sufficient number of circuits as well as circuit pools in advance to train the model.

Fig. \ref{fig:backend} shows how the number of training backends affects reconstruction performance. For both nodes and edges, the average log-ratio mismatch decreases as the model is trained on more backends, indicating improved generalization across hardware. The largest gains occur when increasing the number of backends from one to four. In this range, the model learns backend-agnostic patterns instead of fitting to device-specific behavior. Beyond four backends, the curves flatten, and additional backends provide only modest improvement. Edge mismatches remain higher than node mismatches for small backend counts, reflecting the greater variability of two-qubit operations across devices. With four or more training backends, both node and edge mismatches reach a low and stable level, showing that backend diversity during training is sufficient for accurate error reconstruction. 

\begin{rqsummarybox}
\textbf{RQ2 Summary: }The performance of our proposed framework is most sensitive to data volume and backend diversity: the average log-ratio mismatch drops sharply when increasing circuit pools and training backends, then shows diminishing returns, with edges benefiting the most from added pools/backends due to higher 2Q variability.
\end{rqsummarybox}


\subsection{(RQ3) Comparison to Prior Work}

\begin{table*}[t]
\centering
\caption{Capability comparison between prior forensic ranking and the proposed framework.}
\begin{tabular}{lcc}
\toprule
 & \textbf{Prior Work \cite{roy2025forensicserrorratesquantum}} & \textbf{Proposed Framework} \\
\midrule
Output type & Discrete edge bins / ranks & Continuous node \& edge error map \\
Execution results required & No (ranking); Yes (optimizer) & No \\
Calibration access at inference & Used for evaluation / bounds & Not required \\
Numerical stability & Sensitive to optimizer settings & Stable learning-based inference \\
Circuit scale evaluated & Small circuits (5 qubits, depth 2) & Full device graph, large circuits \\
Cross-backend generalization & Not supported & Supported \\
Scalability & Limited circuits / backends & Full device graph \\
\bottomrule
\label{table:comparison}
\end{tabular}
\end{table*}
We compare our framework with the forensic error analysis proposed by ~\cite{roy2025forensicserrorratesquantum}, as it is one of the closest prior efforts to infer backend error characteristics with user-visible artifacts. Their work introduces two complementary approaches for this task. The first relies on frequency-based edge ranking, while the second derives error rates using an optimizer-based formulation. Table \ref{table:comparison} summarizes the key capability differences between their methods and our proposed framework. These are further elaborated below.

\textbf{Frequency-based edge binning: }
Their first approach ranks qubit links into a fixed number of bins based on how frequently they are selected by the transpiler. This method does not estimate actual error rates and provides only coarse-grained ordering. Even with this coarse grouping, bin assignments are often inaccurate, with only about 83.5\% of links in IBM Sherbrooke and 80\% in IBM Brisbane falling within $\pm 2$ bins of the true rank. Since multiple links within the same bin can have very different error rates, this approach cannot support analyses that require quantitative error estimates. In contrast, our framework predicts continuous error values and achieves more than 80\% accuracy in magnitude.

\textbf{Optimizer-based error extraction: }
The second approach proposed by the prior work \cite{roy2025forensicserrorratesquantum} derives gate error rates by solving fidelity equations using a nonlinear numerical optimizer. This method requires access to circuit execution fidelities, adding extra requirements for forensic analysis. The fidelity expressions are products of multiple $(1-\varepsilon)$ terms, and the observed fidelities tend to be close to one. Since error rates are small, this makes the system weakly sensitive to small changes in $\varepsilon$, leading to a loss of resolution and making it difficult for the optimizer to distinguish nearby error values.

In addition, the use of random circuits causes repeated reuse of the same qubit links, so newly generated equations often introduce limited new information. This results in redundant equations and an ill-conditioned system. As acknowledged by the authors, small differences in measured fidelities can then lead to large swings in the solved error values. Because the error values are small, multiple parameter configurations can satisfy the same fidelity equations, resulting in multiple valid solutions. The optimizer is therefore sensitive to tolerance, initialization, and bounds: tight tolerances increase runtime, while looser tolerances lead to large mismatches. Furthermore, this approach is demonstrated only on small circuits with five qubits and shallow depth, and is not evaluated on large, complex workloads.

In contrast, our framework does not require execution fidelities or nonlinear optimization. It avoids ill-conditioned inverse problems and enables inference across the complete set of physical qubits and couplings.

\begin{rqsummarybox}
\textbf{RQ3 Summary: } Compared to \cite{roy2025forensicserrorratesquantum}, our proposed framework moves from coarse link-ranking and optimizer-based inversion to a stable, learning-based approach that infers a continuous node-and-edge error map at full-device scale and generalizes across backends.
\end{rqsummarybox}

\subsection{(RQ4) Temporal Variation of Noise}
Quantum processors exhibit time-dependent fluctuations in their noise characteristics, which are typically managed through periodic recalibration and generally do not change the average noise levels \cite{temporal}\cite{ghosh_qgan}. Because such drift can influence the forensic procedure, we study its impact using IBM devices that display temporal noise variation. Empirically, we find that the mean deviation in noise parameters for single-qubit gates induced by this variation is on the order of $\sim 10^{-4}$ \cite{mits} and $\sim 10^{-2}$ for two-qubit gates \cite{das2025optimizationquantumerrorcorrecting}\cite{qubit_equal}. Motivated by this observation, we take the calibrated noise models from four of the IBM backends used previously and inject an additional perturbation $\delta_n$ of magnitude $\mathcal{O}(10^{-4})$ for single-qubits and $\mathcal{O}(10^{-2})$ for two-qubit links to emulate realistic temporal drift. During GNN training as well as inferencing, we resample and apply $\delta_n$ every 3 epochs to both the nodes and edges in the hardware backends, thereby capturing the effect of slowly varying hardware noise over time.

\begin{figure}[t]
    \includegraphics[width=0.5\textwidth]{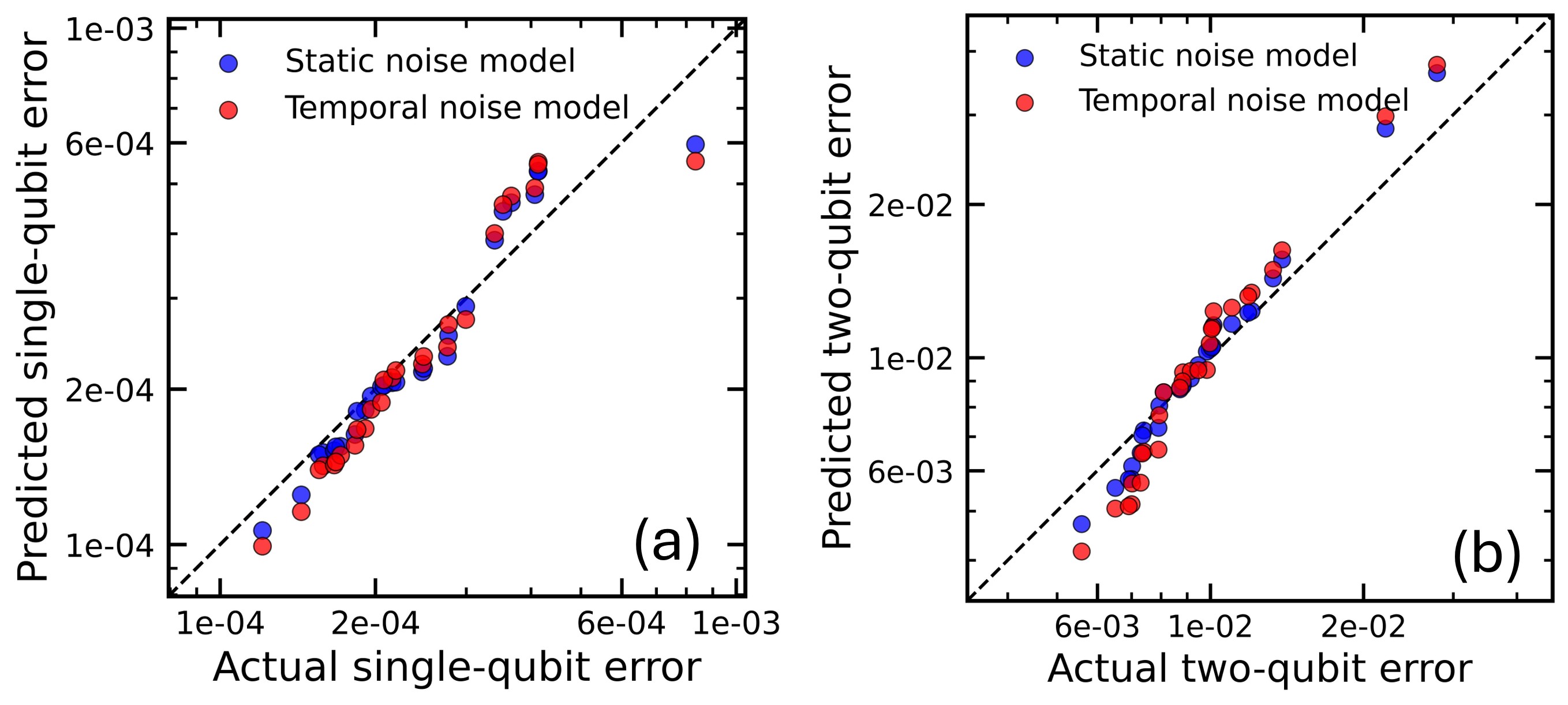}
    \centering
    \caption{Predicted vs. actual (a) node and (b) edge error rates under static and time-varying noise model.}
    \label{fig:temporal}
\end{figure}

Fig. \ref{fig:temporal} shows the robustness of the framework under temporal noise variation. Each panel compares predictions obtained with a static noise model to those obtained when calibrated error rates are perturbed over time.  For both nodes and edges, the predictions under temporal variation remain closely aligned with those from the static model. The overall spread around the diagonal does not increase noticeably, showing that moderate temporal drift does not degrade reconstruction quality. 
The effect is consistent across single-qubit and two-qubit errors, despite the larger absolute scale of edge errors. This indicates that the model captures stable structural patterns in hardware noise rather than 
overfitting to a single calibration snapshot.

Our framework is designed to be robust to temporal noise drift rather than to follow short-term changes in qubit or coupling rankings. It predicts a single representative error value for each component, which reflects average behavior over time. If temporal noise drift causes temporary changes in the relative quality of qubits or links, these changes are not explicitly tracked. In practice, calibration data itself is published as discrete snapshots, so modeling a stable error structure is sufficient for realistic use.
\begin{rqsummarybox}
\textbf{RQ4 Summary: }Even in the presence of a temporal variation of noise, the predictions under the perturbed noise model are closely aligned with the static-noise baseline, indicating that the GNN reconstructs a stable, time-averaged error structure and is not significantly degraded by moderate calibration drift.
\end{rqsummarybox}

\section{Conclusion}
\label{sec:conclusion}
This work presents a data-driven framework for quantum hardware forensics that infers single-qubit and two-qubit error characteristics using only topology information and transpiled circuits. By combining static backend topology with dynamic circuit-level features, the proposed GNN models reconstruct error maps without requiring calibration access to the target device. Extensive evaluation across multiple IBM backends shows accurate error estimation, strong generalization to unseen hardware, and consistent performance under temporal noise drift. 
Across evaluated backends, the framework achieves low average percent mismatch for both single-qubit and qubit-link error rates. The model also preserves strong ranking agreement, with a high correlation between the rankings induced by predicted error values and those induced by the actual calibration errors. Future work will extend the framework to larger-scale quantum processors with higher qubit counts and more complex coupling maps. 

\section{Acknowledgements}
\noindent This work is supported in parts by NSF (CNS-1722557, CNS-2129675, CCF-2210963, CCF-1718474) and Intel’s gift. 

\bibliography{Ref}

\end{document}